# Managing Object-Oriented Integration and Regression Testing (without becoming drowned)

**Mario Winter**
**Dept. of CS**
**Practical Informatics III (Software Engineering)**
**University of Hagen, Germany**
**e-Mail:** `mario.winter@fernuni-hagen.de`

**Abstract:** Systematic testing of object-oriented software turned out to be much more complex than testing conventional software. Especially the highly incremental and iterative development cycle demands both many more changes and partially implemented resp. re-implemented classes. Much more integration and regression testing has to be done to reach stable stages during the development. In this presentation we propose a diagram capturing all possible dependencies/interactions in an object-oriented program. Then we give algorithms and coverage criteria to identify integration resp. regression test strategys and all test cases to be executed after some implementation resp. modification activities. Finally, we summarize some practical experiences and heuristics.

## 1    Introduction

In the last decade of this century object-oriented programming undoubtedly has become one of the mainstream implementation technologies to cope with the ever increasing demands on functionality and quality of software-based solutions. On the quality side, systematic testing of object-oriented software turned out to be much more complex than testing conventional software. As a consequence, productivity gained from OO technology gets partially lost during testing [PerKai90]. To overcome these deficiencies, much work has been published in the last years — most often focusing on either structural or behavioural views of the system[1]. Especially the highly incremental and iterative development cycle proposed by all main object-oriented methodologists [UML98] [Meyer97] demands both many more changes and partially implemented or re-implemented classes. In this presentation, we tackle these challenges with a unifying view on object-oriented integration and regression testing.

The paper is organized as follows. After glancing on notions and previously published results in object-oriented integration and regression testing, in chapter three we take a look on the dangerous chutes object orientation keeps ready in these realms. Then we sketch the required steps for a general regression and integration testing process. In chapter four we focus on code changes, change identification, and the change impact analysis problem, being at the heart of regression testing. We propose a simple diagram capturing all possible dependencies and interactions in a given object-oriented program. In chapters five and six we give algorithms and coverage criteria to identify integration resp. regression test strategys and all test cases which have to be executed after some implementation resp. modification activities. In the last two chapters, we summarize practical experiences and heuristics and give some conclusions.

## 2    Integration and Regression Testing

In waterfall software development process there is a clear distinction between integration and regression testing: *integration testing* is done during development to find errors in the interactions and interfaces of the new, unit-tested modules, while *regression testing* is done during maintenance with the purpose of re-establishing our confidence that the software will continue to function after some modifications. In addition to a big-bang integration of the modules, simple directed strategies like top-down and bottom-up integration testing have been proposed for structured software with tree-like, hierarchical functional dependencies [Spillner90].

Regression testing has to be done both for *progressive changes* (i.e. specification and code changes) and *corrective changes* (code changes only). In the case of progressive changes, because of the changed specification we have to adapt old and to implement new black-box tests. Furthermore we have the possibility of running all our existing tests again, or to try to find out which tests ought to be run again and which ones we don't need to run again. This problem is called *selective regression testing*. Of course we have to be aware of the trade-off — selecting and running the right test cases should be cheaper than running all test cases. But anyway, in the case of progressive changes there may exist test cases which must not be rerun to avoid false alarms.

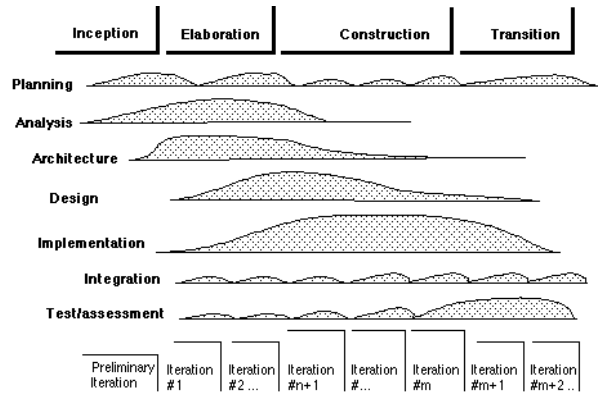

**Figure 1 The Iterative, Incremental Development Process [Kruchten96]**

Regarding object-oriented software, the highly incremental and iterative development cycle blurs much of these distinctions. As seen from a management perspective, i.e., the business and economics point of view, the software lifecycle is organized along four main phases, indicators of the progress of the project [Kruchten96]: *inception*, specifying the end-product vision and its business case; *elaboration*, planning the necessary activities and required resources; specifying the features and designing the architecture; *construction*, building the product and evolving the vision, the architecture, and the plans until the product is ready, and *transition*, delivering the product to its user's, which includes: manufacturing, delivering, training, supporting, maintaining the product until the users are satisfied. Going through the 4 phases is called a development cycle, and it produces a software generation. So software development is now seen as a succession of iterations, through which the software under development evolves incrementally. Each iteration is concluded by the release of an executable product, which may be a subset of the complete product, but useful from some engineering or user perspective. Each release is accompanied by supporting artifacts: plans, release description, user's documentation, plans, etc. An iteration consists of the activities of planning, analysis, design, implementation, and testing in various proportions depending on where the iteration is located in the development cycle (fig. 1). The end of the phases are synchronized with the end of iterations. In other words, each phase is broken down into one or more iterations, such that much more integration and regression testing have to be done, since new and derived classes may have been added, and existing classes may have been modified during an iteration. Because clearly separated phases with cast-iron documents don't exist anymore, many changes are due to a a mixture of progressive and corrective reasons.

In conclusion: to lower the test effort, we are faced with the seemingly different tasks of selecting integration strategies which minimize the need for drivers and stubs, and to precisely isolate the effects of changes on unchanged parts of our system — the so-called change impact analysis task in selective regression testing. Having these inherent difficulties in mind, it is now clear why most of the test budget in object-oriented projects is spent during integration and regression testing [Firesmith95][Boisvert97]. Despite of these realities, only a few publications deal with these problems:

[RothermelHarrold94] proposed change identification and change impact analysis based on control- and data-flow information. [KungEtAl96] also proposed a change identification algorithm based on a combined control- and data-flow representation. Like [Overbeck94] they build their change impact analysis and integration resp. regression testing strategies using the class diagram given by design documents or rebuilt by reverse engineering. [JorEri94] proposed a functional approach to integration testing merely based on the interactions between objects, where simple interactions should be tested before complex interactions involving several objects. At last year's euroSTAR, [Dorman97] explores the problems of testing C++ classes from a practical point of view, and discusses an improved integration testing approach based on an extension to the widely used source code instrumentation technique. Dorman discusses some test metrics measuring the extent to which polymorphic features of the software have been exercised, thus providing more insight into the dynamic complexity of the software under test.

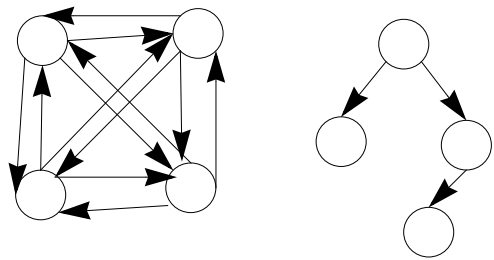

**Figure 2 Complexity: OO vs. Functional**

# 3 The Chutes

In this chapter we present some of the chutes lurking in the broad river of object-oriented integration and regression testing, which often cause testers and managers getting drowned by their first object oriented projects.

## 3.1 Chute 1: Complexity

Object-oriented software systems are composed of objects. What is an object? [UML98] answers

> *"An entity with a well-defined boundary and identity that encapsulates state and behaviour. State is represented by attributes and relationships, behaviour is represented by operations, methods, and state machines. An object is an instance of a class."*

Objects communicate and influence each other through state changes triggered by messages. If and how an object reacts on a message is determined by its state, as are the objects visible to a certain object at a certain point in time. So objects at runtime build a time-variant network of communicating entities having multiple entry points and being created and deleted dynamically, responding to triggering events. As network theory shows, n objects (or the developers implementing the corresponding classes) can interact through $n^2$-n channels at most. In comparison with functional decomposed designs (i.e. call trees), with only n-1 interfaces for n functions, this quadratic increase of interactions may bury our efforts in the complexity tarpit (fig. 2). Together with uncertainties of polymorphic interactions, we appreciate good layered architectures with high cohesion and low coupling.

## 3.2 Chute 2: Cyclic dependencies

Even in sophisticated designs, as captured through design patterns [Gof94], we find mutual dependencies. For example let's look at the Mediator pattern, the structure of which is shown in fig. 3. The intent of mediator is to

> *... define an object that encapsulates how a set of objects interact. Mediator promotes loose coupling by keeping objects from referring to each other explicitly, and it lets you vary their interaction independently.[Gof94]*

As we see, cyclic dependencies are introduced between classes **ConcreteMediator** and **ConcreteColleague1** resp. **ConcreteColleague2**. Given that design patterns capture best practices in object-oriented design, as a corollary integration strategys like top-down or bottom-up are meaningless, because testers are confronted with the dilemma

> *"How should one integrate and (re-) test a system with no top?"*

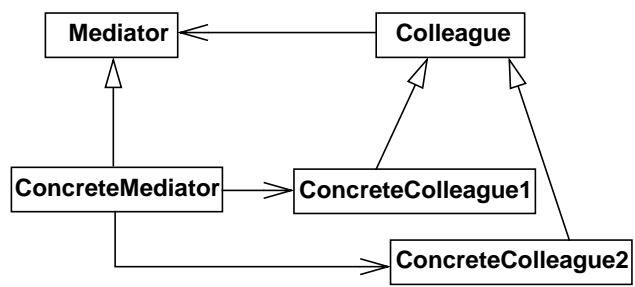

**Figure 3 Mediator design pattern**

### 3.3   Chute 3: Cost, or Knowing when to Stop

Let us consider we are living in a brave world and test cases exist for our system. In case of interactive applications we will have many GUI test cases, the replay of which is time consuming. Full regression testing may be prohibited due to short build cycles — as an example, Microsoft's developers in most projects synchronize each day, with testing done at night [CusumanoSelby95]. In consequence, one has to select a minimal set of revealing test cases, and to skip those which are not affected by the changes in the last iteration. In addition adequate integration coverage criteria are needed because sufficient tested classes don't imply a sufficient tested system — the whole is more than the sum of its parts. Even if we know all dependencies and adequate coverage criteria, we may lack time to re-run all concerned tests. One common approach is "Stop when the budget says so", but we should do better.

### 3.4   Does a Life-Jacket Exist?

In despite of these object-oriented chutes there is a good message: at a second glance we observe that the addition of a new base class to the system is a special case of adding a subclass (integration testing without superclass), which in turn is a special case of modifying an existing class (regression testing). Regarding this unification, in object-oriented software development we should be able to use one technique for both regression and integration testing. In fact we can sketch a general process of integration and regression testing during object-oriented soft-

```
FOR each increment DO
        change identification;
        change impact analysis;
        generate (re-) test strategy;
        select/generate black box test cases;
        execute black box test cases;
        UNTIL covered DO
                select/generate white box test cases;
                execute white-box test cases;
        END
END
```

**Figure 4 General Integration and Regression Test Process**

ware development (fig. 4). In the sequel we look at the steps of this process in detail.

## 4   Code Changes

According to [KungEtAl96] we classify changes based on the granularity of the changed element (variable declaration, statement, method, class, cluster) and the special type of the change (addition, deletion, modification). To illustrate and to cope with these types of changes we introduce the class message diagram, a new diagram for object-oriented software combining the behavioural and structural aspects influencing integration and regression testing.

### 4.1   The Class Message Diagram

Let us first look at messages, being the basic vehicle of interaction in object-oriented programs. We observe that each message emerges from a unique, syntactically identifiable position in a particular method defined in a particular class. The reception of the message by an object leads to the execution of a particular method, which only can be identified only by the class of the receiving object — due to polymorphism this could be any class in the case of untyped object-oriented languages and any subclass of a particular, syntactically identifiable class in the case of static typed object-oriented languages.

Since we focus on integration and regression testing, we abstract from control- and data-flow analysis, but merely look at possible interactions, i.e. the interfaces and "call-" or "uses-" relations between methods. Suppressing inheritance and polymorphism for the moment, we define the *class message diagram* (CMD) to be composed of *method nodes* representing methods. A directed *message edge* connecting method[2] a in class **A** with the method b in class **B** means "during the execution of method a in an object of class **A** a message could be sent to an object of class **B**, leading to the execution of method b". In other words, method a defined in class **A** uses method b defined in class **B.**

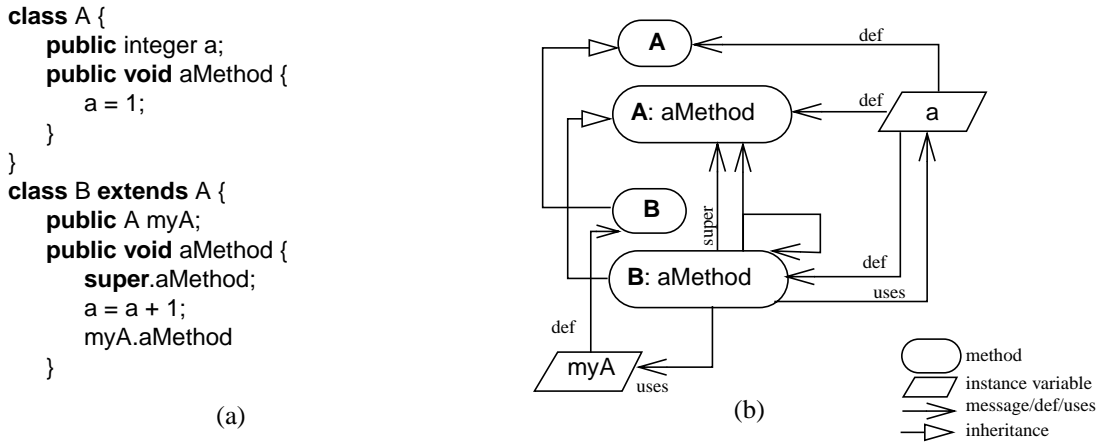

```
class A {
    public integer a;
    public void aMethod {
        a = 1;
    }
}
class B extends A {
    public A myA;
    public void aMethod {
        super.aMethod;
        a = a + 1;
        myA.aMethod
    }
```

(a)                                          (b)

**Figure 5 Sample Java Code and its Class Message Diagram**

Now we consider inheritance, polymorphism, and the accompanying **self** and **super** method selection constructs within the CMD. Each occurrence of **self** in a method defined in some class leads to a single message edge labelled "self" to the first implementation of method a upwards the inheritance hierarchy of (and *in*cluding) this class. In the same manner each occurrence of **super** a in some method defined in some class leads to a single message edge labelled "super" to the first implementation of method a upwards the inheritance hierarchy of (and *ex*cluding) the class. We add an *inheritance edge* from method a in class **B** to method a in class **A** if class **B** is a subclass of class **A**, i.e. method a in class **B** redefines method a of class **A**. In this case all unlabelled message edges ending at method a are duplicated, ending at method a in class **B**. These duplicated and redirected message edges reflect the fact that "calls" of method a could be polymorphically dispatched to instances either of class **A** or class **B**.

Note that the CMD is defined on the class level but covers aspects occurring at the object level, like aliases and actual orderings of messages, though we abstract from modelling intra-method control flow dependencies. This stems from the broadcast-based system model underlying the CMD, where we may think of all messages beeing sent simultaneously to all instances, thus building an upper bound for the dependencies possibly occurring at "runtime". For finer analyses we may also look at instance variables capturing the state of objects and add a *data node* for each instance variable to the CMD. Regarding accesses to variables as special messages, an edge from a method node a to an instance variable node x labelled "uses" means "during the execution of method a instance variable x is accessed", an edge from an instance variable node x to a method node a labelled "def" means "instance variable x may be modified by the execution of method a".

Lets look at a simple example. Consider the code fragment in Java shown in fig. 5 (a). The CMD for classes **A** and **B** is shown in fig. 5 (b). First of all we see four method nodes rather than two, because default constructors for both classes are included. Note that def edges connect the default constructor method with all instance variables declared in a class. An inheritance edge denotes the redefinition of aMethod by class **B**, the message edges from **B**'s aMethod to aMethod in class **A** reflect the usage of **super** in class **B**'s aMethod and the (polymorphic) usage of myA.aMethod. The latter leads to the self-recursive message edge of **B**'s aMethod, too. The def edges to both aMethod nodes and the uses edge from aMethod in class **B** are self describing.

## 4.2   Safe Change Impact Analysis

After an iteration we identify source code changes in a three level process:

1. Identify all changed (added, modified, or deleted) classes and methods (with the help of our configuration management system or, in the worst case, relying on directories, file dates and versions, and textual comparisons.)

2. Identify all changed data definitions (textual, by parsing, or through a data dictionary).

3. Given the CMD's before and after the changes, mark all changed nodes and edges in the CMD.

Now we are ready for the next step of our integration and regression test process — the change impact analysis. Surely an algorithm for selective regression testing must regain its own costs through the profit of not running the dropped test cases. To cope with this balance, a battery of change impact analysis methods are published, where

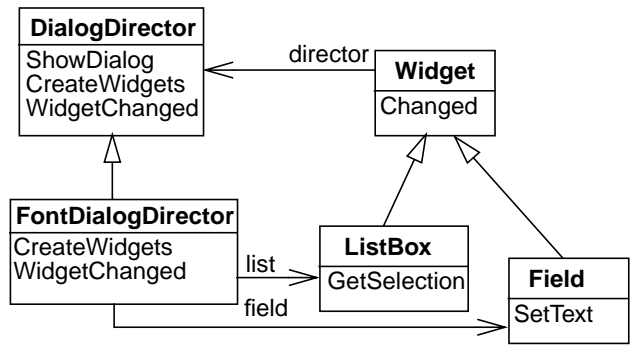

**Figure 6 Mediator: DialogDirector**

*"Impact analysis is the act of identifying the potential effects of a change, or estimating what needs to be modified to accomplish a change." [BohnerArnold96].*

We call an impact analysis *precise* if it doesn't select software entities not concerned by a change. An impact analysis is called *save* if all concerned software entities are included in the selected ones. Surely precise techniques are more expensive than safe techniques. Dataflow- and slicing-based techniques for object-oriented software, which promise to be both precise and safe are still in their infancies [RothermelHarrold94], so most code-based change impact analysis techniques for object-oriented software examine transitive dependencies, i.e. they compute the transitive closure of the underlying dependency relation[4] [AbdullahWhite97] [KungEtAl96] [Mohl97].

Change impact analysis based on the CMD is done in four steps. Firstly we remove all inheritance edges from the CMD (remember that polymorphism is considered by duplicating and redirecting corresponding message edges). Secondly we collapse all parallel edges, thus converting the multigraph structure of our CMD to a simple, directed graph. Thirdly we transpose the graph, i.e. we reverse the direction of all edges (because we search for all methods depending on and all variables defined by the changed method). In the fourth step we do a depth-first search on the resulting structure.

Consider our mediator design pattern example, an actual instance of which is shown in fig. 6 [Gof94]. Let us assume that method WidgetChanged in class **FontDialogDirector** is modified. Change impact analysis based on the class diagram reveals that all five classes have to be re-tested, since they form a strong component (i.e. they are mutual reachable via the defined or inherited associations). To cope with this problem we build the CMD (left side of fig. 7, def/uses edges omitted), thus changing the granularity of our observations to the method level. We note that both cycles present in the corresponding class diagram (fig. 6) are broken, meaning that some associations are not traversed by the messages emerging from method WidgetChanged. In our example the CMD based change impact analysis reveals that in addition to the originally changed method WidgetChanged only method Changed defined in class **Widget** has to be re-tested, a test effort reduction of some 90%.

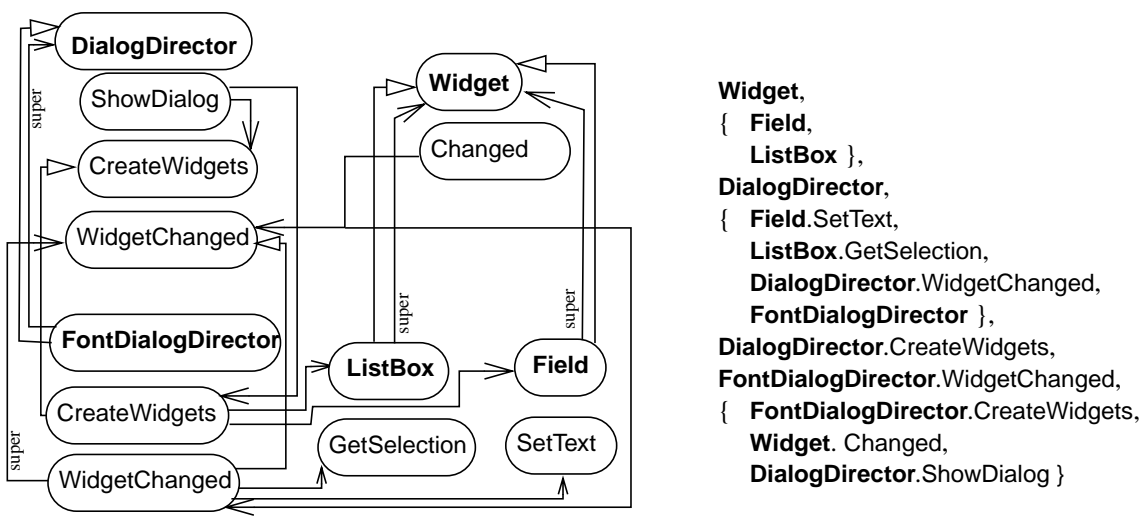

**Figure 7 DialogDirector Class-Message Diagram and Test Strategy**

# 5 Test Selection and Strategy Generation

To select tests to be re-run we need traceability from test cases to requirements, design and code. This could be realized by the use case and test scenario approach shown in [KöPaRiWi97] together with appropriate versioning schemes and configuration management. After identifying the concerned methods we have to select all test cases examining these methods. To accomplish this task together with the actual outcomes of a test we have to save the trace, i.e. the methods executed and messages sent by running the test. Simple method instrumentation suffices, since the information of when a method is started and ended reveals the whole trace.

Given the concerned methods and test-cases resp. all (new) methods for integration testing, with the aid of the CMD with inheritance edges we can generate a test strategy in two steps:

1. Condense all strong components to one artificial node[5].

2. Compute the topological order of the resulting nodes.

This results in a lattice of methods and strong components of methods, on which we can adapt strategies like bottom-up, top-down, or outside-in. Methods on the same level in the resulting partial order may be integrated and tested in any order, but touched test-cases derived from unchanged specifications should be run first. Clearly constructors should be tested before testing any other methods of the class. Strong components on the method level are broken by stubbing some methods to break the cycles.

Remember that in our mediator example both cycles formed by the director, list, and field associations are broken by the CMD. So with the aid of the CMD we immediately are able to compute the test strategy depicted on the right side of fig. 7. According to the results of [PerKai90], methods CreateWidgets and WidgetChanged of class **DialogDirector** are tested before their corresponding redefining methods in class **FontDialogDirector**, thus enabling the reuse of test cases and results.

# 6 Coverage Criteria

Based on the CMD we define the following interaction based coverage criteria for object-oriented integration and regression testing:

- *method coverage* ($c_o$-0): all method nodes of the class message graph have to be visited at least once, i.e. each method in the system has to be executed at least once to achieve 100% method coverage.

- *message coverage* ($c_o$-1): we partition the unlabelled message edges of the CMD due to their message source. Covering one edge from each of the resulting equivalence classes assures that each each message source has „fired" at least once.

- *polymorphic message coverage*: each message edge of the class message graph is covered at least once, i.e. each message source has „fired" at least once for each possible polymorphic binding.

- *system boundary interior test*: polymorphic message coverage with the additional condition, that all cycles in the CMD have to be covered 0, 1, and more than one times.

- *polymorphic complete message coverage* ($c_o$-n): all path's in the CMD have to be covered.

In the case of pure, non-hybrid object-oriented languages like Smalltalk-80, message coverage should be regarded as the minimal criterion for the message-based dynamic integration test of object-oriented software, if branch coverage is guaranteed for each method by appropriate class tests. If no class tests have been done, we should achieve polymorphic message coverage, since for pure object-oriented languages it subsumes statement coverage. Message coverage and polymorphic message coverage correspond to the context coverage metrics proposed by [Dorman97]. As its structured counterpart, $c_o$-n-coverage is achievable only for cycle free CMDs. In addition, we could define data-dependency related coverage criteria based on the CMD. Since such criteria rely on further data-flow analyses, we did not include them into this paper, but refer to [HaworthRoper97].

If time doesn't allow the execution of all test cases, we propose risk analysis for black-box test cases based on the specification raising the test, the criticality of the tested parts (trace), and the depth of the trace of each test case.

# 7 Experiments

We implemented a prototypical test environment for VisualWorks Smalltalk-80 and experimented with the VisualWorks class library and some rather small programs [Mohl]. Smalltalk-80 as an untyped language represents the worst case for testers, since static analysis is very tough so that reliable and fast type inference algorithms for Smalltalk are not available today. A deteriorated result for inheritance hierarchies of library classes (class **Collection** and 29 of its subclasses) is shown in table 1.

| | Class Diagram | Class Message Diagram |
|---|---|---|
| Nodes | 30 | 426 |
| Edges | 395 | 3325 |
| Max in-degree | 30 | 335 |
| Strong components | 1 | 3 |
| Classes in strg. components | 26 | 22 |
| Methods in strg. components | 424 | 157 |

**Table 1. Experiment: Smalltalk Container Classes**

As we see in the first row of table 1, we scrutinized 30 classes resp. 426 methods. The second row reveals 395 (out of a theoretically maximum of 870) associations between the 30 classes (inferred from the message selectors) and 426 (out of a theoretically maximum of 181,050) message edges between methods. The maximum in-degree of a class (i.e. the number of directly dependent classes resp. methods) is 30, where in case of methods it is 335. The last three rows indicate one strong component on the class level including 26 out of the 30 (!) classes and 424 methods, which on the method level decays into 3 strong components including 22 classes, but only 157 methods. Clearly the high coupling of the classes via the deep inheritance relation faces the tester with big problems.

A somewhat more typical experiment was based on a simple program simulating a vending machine. In this example the inheritance hierarchy is rather flat and the results are more promising for testers. Especially the strong component formed by two of the classes decays to several linear calling sequences on the method level. In this case we are able to test totally without stubbing. The lessons learned by these experiments is twofold: we can't test-in quality — but well designed programs make testing much easier.

| | Class Diagram | Class Message Diagram |
|---|---|---|
| Nodes | 6 | 30 |
| Edges | 12 | 36 |
| Max out-degree | 5 | 4 |
| Strong components | 1 | 0 |
| Classes in strg. components | 2 | 0 |
| Methods in strg. components | 19 | 0 |

**Table 2. Experiment: Vending Machine**

# 8   Conclusions

## 8.1   Pragmatics and Lessons Learned

Due to the dramatically increasing communication overhead, incremental and iterative development is impossible without good architecture. So we grow up our systems framed in an architecture of subsystems and building-boxes, allowing our development teams to work rather isolated, relying on small interfaces to other teams. As an example, in the sense of good object-oriented design we argue that instance variables are only being read or modified by set- resp. get-methods defined within the class declaring the instance variable (or its subclasses), thus breaking cycles involving def/use edges in the class message diagram. Subclasses should not change the semantics of any inherited attributes or methods, but may add attributes and/or define or redefine methods. We build and compare the CMD against design documents to check up whether or not architecture violations are lurking in our code.

We emphasize on integration and regression testing for object-oriented programs using strategies combining structural and behavioural aspects. Our approach is based on three observations:

1. There is no real difference in object-oriented regression and integration testing.

2. Cyclic dependencies occurring at the class structure level most often decay into several smaller components or even into linear call sequences at the method-message level.

3. Pragmatical change impact identification and integration resp. regression strategy generation based on the CMD is sufficient for well designed object-oriented programs.

When we automate our test suites, testware becomes software. So rather than capturing lots and lots of GUI-inter-actions, we plan and implement modular test-suites based on scripting or programming languages. We trace tests to the application and keep tests under configuration managment, too.

We trace requirements like use cases to be implemented in the next iteration down to the subsystems, classes, and methods which have to be modified. Together with change impact analysis algorithms integrated as skripts in our case tool we are able to answer "what if" questions like

- risk analysis for proposed changes,

- planning of changes,

- tracing the effects of changes.

Moreover, we use change impact analysis for understanding and documenting software.

We currently are porting our Smalltalk-80 test environment to Java, where static typing brings much more pre-ciseness to our pragmatic approach. Using the reflectional facilities introduced in jdk 1.1, we are able to choose at runtime between instrumented and original classes to make testing even more convenient

## 8.2 Acknowledgements

We thank Gabriele Mohl for her prototypical implementation of the presented ideas and her experiments in the VisualWorks Smalltalk-80 environment.

# 9 Footnotes and References